\documentstyle[epsf,elsart12]{article}

\begin{document}
\baselineskip=16.0pt

\begin{center}
{\Large \bf Cluster Analysis of Extremely High Energy }

{\Large \bf Cosmic Rays in the Northern Sky}

\bigskip
        Y.Uchihori$^1$,
        M.Nagano$^2$,
        M.Takeda$^3$,
        M.Teshima$^3$,
        J.Lloyd-Evans$^4$,
   and  A.A.Watson$^4$
\end{center}

\medskip
\noindent
{\it\small$^{1}$  International Space Radiation Laboratory, 
National Institute of Radiological Sciences, \\
Chiba 263-8555, Japan} \\
{\it\small$^{2}$  Department of Applied Physics and Chemistry,
Fukui University of Technology, Fukui 910-8505, Japan} \\
{\it\small$^{3}$  Institute for Cosmic Ray Research, University of Tokyo,
    Tokyo 188-8502, Japan} \\
{\it\small$^{4}$  Department of Physics and Astronomy, University of
Leeds,
    LS2 9JT Leeds,
UK} \\

\bigskip

\begin{abstract}
The arrival directions of extremely
 high energy cosmic rays (EHECR) above $4\times10^{19}$ eV, observed
by four surface array experiments in the northern hemisphere,
are examined for coincidences from similar directions in the sky.
The total number of cosmic rays is 92.
A significant number of double coincidences (doublet)
and triple coincidences (triplet)
are observed on the supergalactic plane
within the experimental angular resolution.
The chance probability of such multiplets from a uniform distribution is
less than 1 \%
%a few $\times 10^{-3}$,
if we consider a restricted region
within $\pm 10^{\circ}$ of the supergalactic plane.
Though there is still a possibility of chance coincidence, 
the present results on small angle clustering
along the supergalactic plane may be important
in interpreting EHECR enigma.
An independent set of data is required to check our claims.  \\

 extragalactic)),
  98.65.-r (Galaxy groups, clusters, and superclusters; large scale
 structure of the Universe)

\end{abstract}

\normalsize

\section{Introduction}
It is well known that extremely high energy cosmic rays (EHECR) are
subject to photopion production by interactions with primordial
photons on traversal through intergalactic space,
attenuating their energies down to 4$\times10^{19}$eV (40EeV)
if their sources are distributed uniformly over cosmological distances.
In this case the arrival directions of the highest energy cosmic rays
around 40EeV are expected to be almost isotropic and some of them 
above 40EeV may be
correlated to the topological structure of nearby
galaxies since the distance to their birth places
is no further than 50Mpc and the intergalactic magnetic
field is at most 10$^{-9}$Gauss \cite{rachen93}.
If powerful radio galaxies are sources of EHECR's
a correlation of their arrival directions
on the sky with the supergalactic (SG) plane might be expected,
as suggested by
Biermann and Stanev \cite{biermann95}, since extragalactic radio sources
concentrate towards the SG plane and this concentration extends
to at least z $\sim$ 0.02, based on the MRC Catalog measured
at Molonglo covering the declination band between 18.5$^{\circ}$
and -18.5$^{\circ}$ \cite{shaver89}.
In 1995 Stanev et al. \cite{stanev95} studied the arrival direction
of cosmic rays with energies greater than 40EeV and found that they
exhibit a good correlation with the direction of the SG plane. 
The magnitude of
the observed excess is 2.5$\sim2.8\sigma$, in terms of Gaussian
probabilities.  
They used 42 events of the world data published
at that time with 27 events being from the Haverah Park experiment.

The AGASA group claimed  \cite{hayashida96} that
three doublets of showers
with angular separation of less than 2.5$^{\circ}$ 
(consistent with the experimental resolution)
are observed among the 36 events above 40EeV,
corresponding to a chance probability of 2.9\% from
a uniform distribution, but noted that
a significant fraction of EHECR's are uniformly
distributed over the observable sky.
Two of three doublets are
observed within 2.0$^{\circ}$ of the SG plane.
Adding the AGASA data up to August, 1998 and reevaluating
the energies and arrival directions of all AGASA events,
Takeda et al. \cite{takeda99} found one triplet and three doublets
in a total of 47 events.
The chance probability is smaller than 1 \% and the significance
increased with the increased data set.
It is interesting to note that the centroids of the triplet and 
one of the doublets are 0.7 and 0.9 degrees off
the SG plane respectively.

In the northern hemisphere, four surface array experiments
at Volcano Ranch (VR), Haverah Park (HP), Yakutsk (YK) and Akeno (AG)
have been operational so far and, with 81 events above 40EeV, 
we have reported in \cite{uchihori96}
that a significant number of doublets and triplet are observed
around the SG plane with a chance probability
less than 1 \% supporting the
AGASA result.

It is important now to examine whether the significance
of such a clustering of EHECR's in the sky in correlation with
the SG plane increases or not when using the world data set
which includes the new published AGASA data.
If clusters along the SG plane are real, source models which
do not relate to the SG plane may find difficulty in explaining
the observations.

In this report we examine the arrival direction distribution
of the events so far published by the four experiments in the
northern hemisphere.
While we report rather low probabilities of various occurrences
we emphasis that we regard these claims as charting the way
for the analysis of future, independent data sets rather than
as providing conclusive evidence of anisotropy.

%\section{Experimental Data}

\section{Analysis and results}

The experimental data reported by VR \cite{exp_VR},
HP \cite{stanev95}, YK \cite{afanasiev96} and AG \cite{takeda99} are
used in the following analysis.  
The Fly's Eye events
have differing error ellipses, event by event \cite{uchihori96},
which makes the estimation of the chance probability complicated
and the exposure in right ascension of the Fly's Eye instrument is
less uniform than that of the ground arrays.
Hence they are not included in the present analysis.

Experimental details for the four experiments and conditions
for the present analysis are summarized in Table \ref{tab:site}.
Only extensive air showers (EAS) with zenith angles within
45$^{\circ}$ are used in the present analysis
to select EAS of good quality in energy and 
arrival direction determination.
The arrival directions are determined by the arrival time of
the shower front at different detectors,
resulting in the estimated directional uncertainties
evaluated by each experiment
as listed in Table \ref{tab:site} for 40EeV events.

The method of energy estimation is different in each experiment.
In this report we use 92 extensive air showers whose energies are
estimated to be above 40EeV by each experiment and do not normalize
the energy between experiments.

%table1

\begin{figure}[H]
\centerline{\epsfile{file=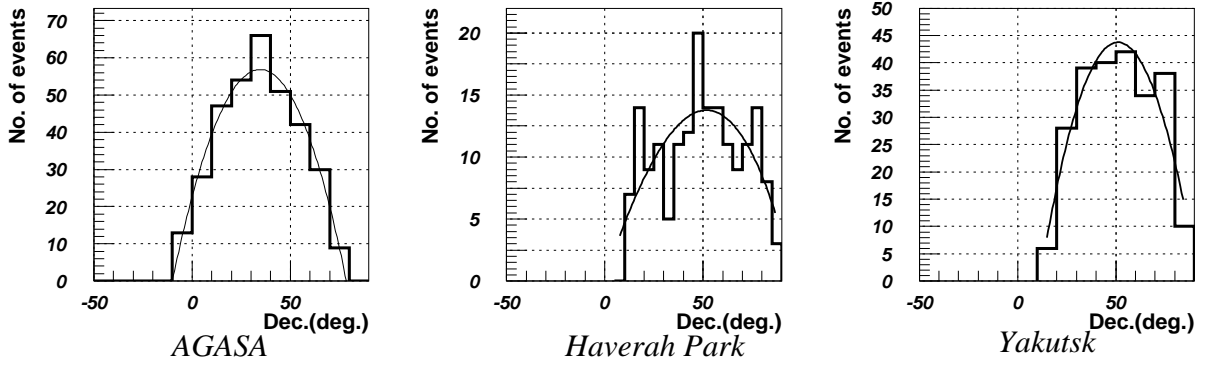,height=9cm}}
\caption{\it  Declination distribution of showers of energies above 10EeV
      and zenith angles smaller than 45$^{\circ}$ of AG, YK and HP.}
\label{fig:decl_GA}
\end{figure}

The observable sky is different for each experiment and is not uniform
in declination.
In the case of the ground array experiments listed
in Table \ref{tab:site}, the right ascension distribution is almost
uniform.
The declination distributions for observed events above 10EeV is,
however different for each experiment as it depends upon the latitude
of the array and
detector type, as shown in Figure \ref{fig:decl_GA}.
Since the triggering efficiency of each experiment becomes uniform
over their array area above 10EeV, the declination
distribution of observed events above
40EeV in each experiment is similar to that above 10EeV.

%fig1,2

\subsection{Galactic and supergalactic latitude distribution}

\begin{figure}[H]
\centerline{\epsfile{file=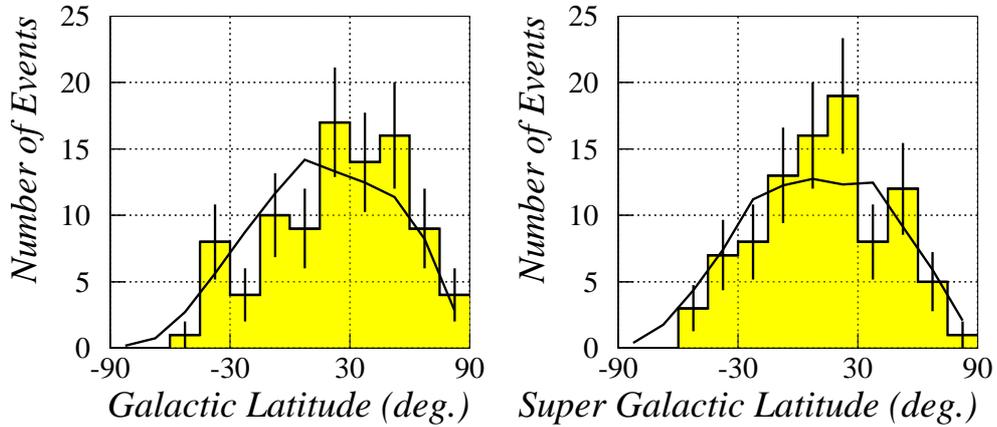,height=7cm}}
\caption{\it   Galactic (left) and supergalactic latitude (right)
      distribution of arrival directions of 92 cosmic rays
      from four ground array
      experiments.
      Energy thresholds are 40EeV.   Solid lines are the sum of expected
      distributions of each experiment for a uniform distribution.}
\label{distribution}
\end{figure}

To study the general arrival direction distribution of 92
events,
the latitude distribution in galactic (G) and SG coordinates are
shown in Figure \ref{distribution}.
Solid lines show the sum of the expected distributions assuming a uniform
incident arrival direction distribution for each experiment;
a uniform distribution in right ascension and the 
observed declination distributions of Figure \ref{fig:decl_GA} 
are used to obtain these curves.
There are no statistically significant deviations from uniformity
in G coordinates.
In the SG coordinate system, there is some excess 
between $0^{\circ}$ and $30^{\circ}$,
but it is not statistically
significant.
Most EHECR's, then, are found to be uniformly distributed over
the observable sky.

%fig3

\subsection{Clustering of arrival directions of showers}

\begin{figure}[H]
\centerline{\epsfile{file=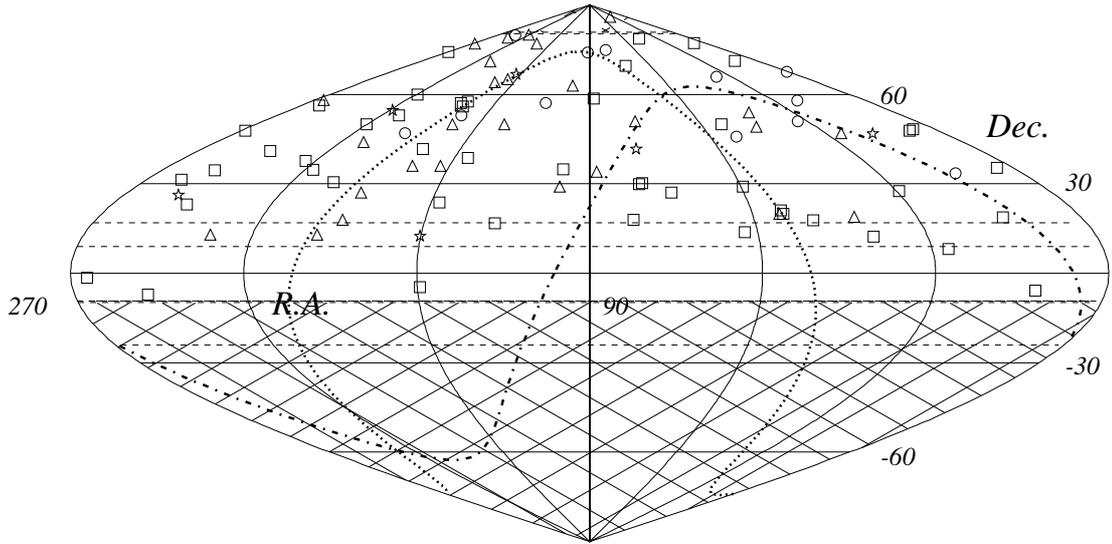,height=8cm}}
\caption{\it Sky map of 92 events above 40EeV in equatorial coordinates.
 Squares - AGASA, Triangles - Haverah Park,
 Circles - Yakutsk and Stars - Volcano Ranch.
The region of sky observable in each experiment is shown by dashed lines
from the top YK, HP, AG and VR.}
\label{sky_map_eqc}
\end{figure}

\begin{figure}[H]
\centerline{\epsfile{file=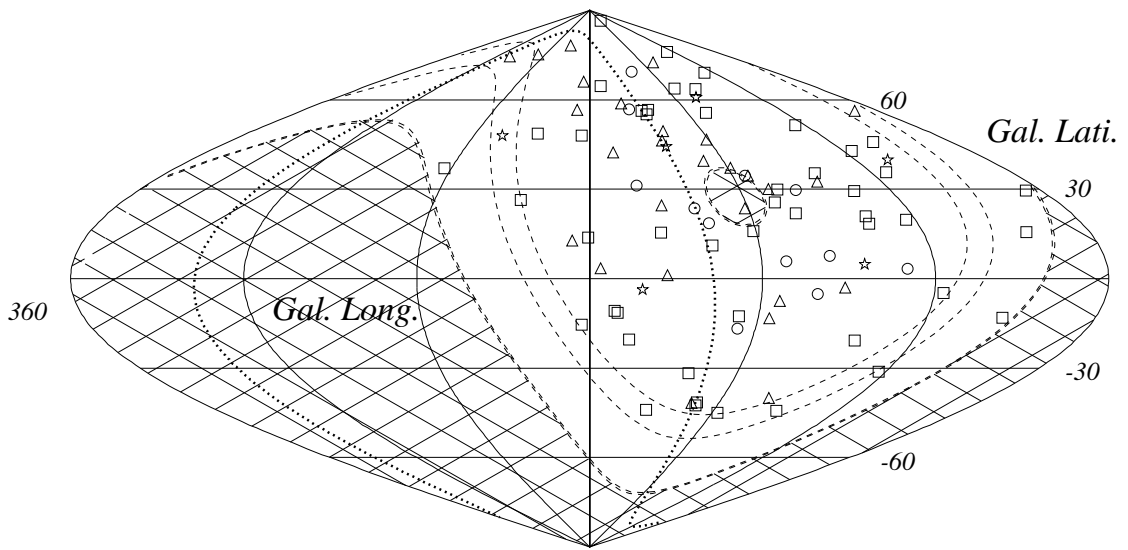,height=8cm}}
\caption{\it Sky map of 92 events above 40EeV in galactic coordinates.
             The symbols are the same as used in figure 3.}
\label{sky_map_glx}
\end{figure}

In Figures \ref{sky_map_eqc} and \ref{sky_map_glx}, the arrival directions
of each event from the four arrays are plotted
in equatorial and galactic coordinates, respectively.
Some coincidences of events coming from similar direction are apparent.
In the following we estimate the chance probability of
such coincidences arising from a uniform incident arrival direction
distribution.
An estimate of the space angle scatter due to errors in 
the arrival direction determination is obtained by quadrature
addition of the uncertainties of Table \ref{tab:error} and
ranges from $2.5^{\circ}\sim 4.2^{\circ}$,
depending on the relevant combination of experimental uncertainties.
Since 80 \% of all events are from AGASA and Haverah Park,
it is sufficient to examine clusters within
$3^{\circ}\sim 4^{\circ}$ for combined data set.

%fig4,5

%table2

The number of triplets and doublets
 within space angles of $3^{\circ}$, $4^{\circ}$ and
 $5^{\circ}$ are listed in  Table \ref{tab:prob}.
In counting doublets in this table, each triplet is also decomposed 
into 3 doublets corresponding to the possible pair combinations 
with differing space angle separations.
So there are actually 2 triplets and 6 independent doublets
within a space angle $3^{\circ}$, but the 
2 triplets are also counted as 6 doublets, making a total of 
12 doublets as listed in the
Table.
To estimate the probability of obtaining such clusters by chance 
from a uniform distribution,
we simulated the same number of events from each experiment.
1,000,000 data sets, each comprising 92 showers,
were simulated under the following assumptions :

\begin{enumerate}
\item The declination distributions of AG, HP and YK events
      are approximated by smooth functions as shown by the solid lines
      in Figure \ref{fig:decl_GA}.
      The AG declination distribution
      is used to simulate the VR data set.

\item A uniform distribution in right ascension with the
      observed declination distributions is assumed for all
      experiments.

\end{enumerate}

\begin{figure}[H]
  \centerline{\epsfile{file=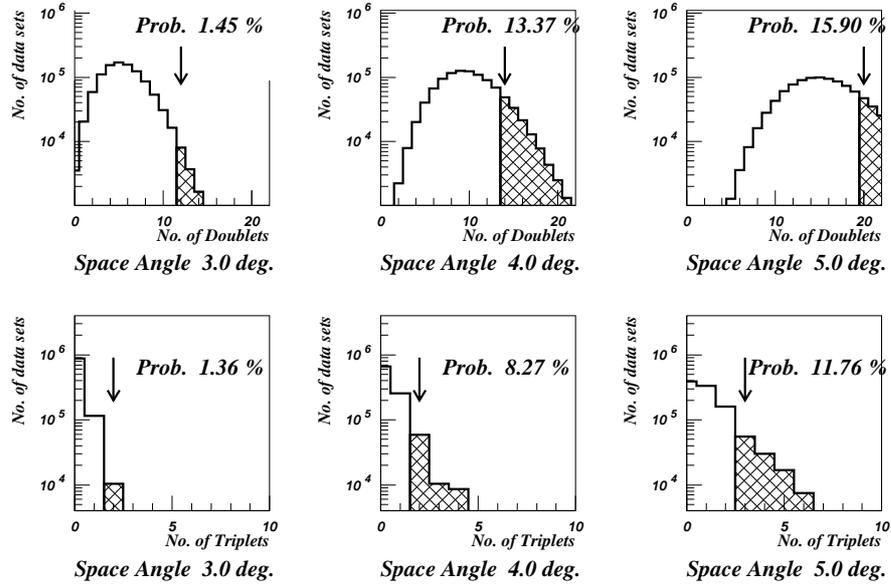,height=7cm}}
  \caption{\it Frequency distribution of the number of doublets in 1,000,000
               simulated data sets within several space angles.
               The arrow mark shows the observed number of doublets.
               Chance probability of
               the observed number of doublets is calculated by summing
               the number of data sets in the hatched region.}
  \label{fig:prob_clust}
\end{figure}

In Figure \ref{fig:prob_clust}
the frequency distributions of the number of doublets found by simulation
of 1,000,000 data sets
within three different space angles separations are shown.
The arrow shows the observed number of doublets.
The chance probability of observing doublets is calculated
as that fraction of  the simulated data sets which have equal, 
or more doublets than observed.

%fig6

The chance probabilities of observing triplets or doublets within the
three space angles separations are listed in Table \ref{tab:prob}.
The probability is small only for clustering  within 3$^{\circ}$.
The chance probabilities
for clustering within 4$^{\circ}$ and 5$^{\circ}$, are larger 
than 10 \% and hence 
not statistically significant.

%table3

\subsection{Correlation with the Supergalactic Plane}

\begin{figure}[H]
\centerline{\epsfile{file=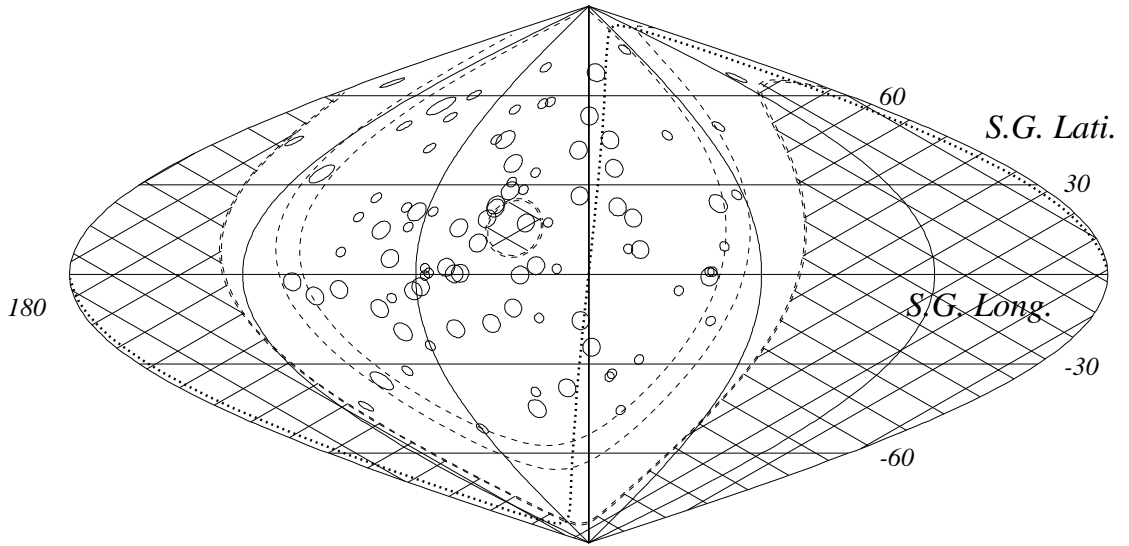,height=8cm}}
  \caption{\it Arrival directions with error circles of EHECR
               in supergalactic coordinates.
               The total number of events is 92.}
  \label{fig:skyplot_sg}
\end{figure}

The arrival directions of EHECR of the four ground array experiments 
are plotted
in the SG coordinate system in Figure \ref{fig:skyplot_sg}. 
Error circles are also shown; 
$1.8^{\circ}$ in the case of AG, and $3.0^{\circ}$ for HP, VR and YK.

%fig7

\begin{figure}[H]
  \centerline{\epsfile{file=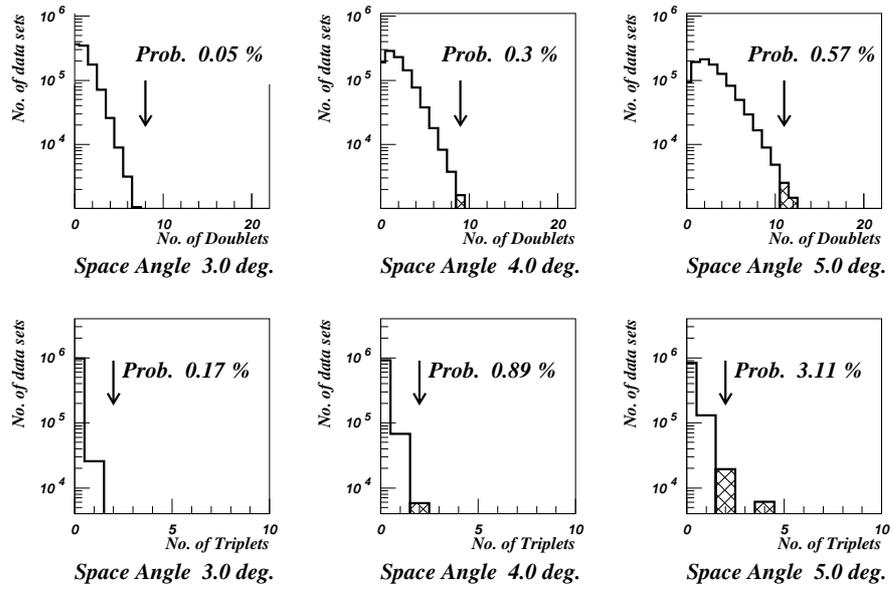,height=7cm}}
  \caption{\it Probability of observing doublets and triplets within
               10$^{\circ}$ from the SG plane with 1,000,000 simulated
               data sets for 92 events.}
  \label{fig:prob_sg}
\end{figure}

Note that the two centroid of triplets are in a very narrow region
(within 0.9 degrees) about the SG plane.
An estimate of the chance probabilities of obtaining this
number of doublets within $\pm 10^{\circ}$
from the SG plane is shown in Figure \ref{fig:prob_sg}.
In this analysis, only the number of doublets within $\pm 10^{\circ}$ region
is counted and compared between observed data and simulated data,
without taking into account doublets outside this region.
In Table \ref{tab:prob_sg} the chance probabilities of doublets and triplets
within $\pm 10^{\circ}$ from the SG plane are listed.
Whilst the choice of a cut at 10 degrees about the SG plane is
arbitrary, we note that the chance probabilities for doublets 
within all space angle separations, and for triplets within the two 
smaller space angle cuts, are intriguingly below 1 \%.

%fig8

%table4

\section{Discussion}

As Figure \ref{distribution} indicates, there is about a 20 \% excess
of events in the latitude band $0 \sim 30^{\circ}$ about the SG plane.
Though this excess is only a one $\sigma$ effect, it may 
contribute to the apparently significant clustering 
estimated above, so 
we have also calculated the chance probability of multiplets,
given a uniform distribution within this band but with a total 
event number excess of 20 \%.
The result
is  listed in Table \ref{tab:prob_excess}.
As expected the statistical evidence for clustering is reduced 
somewhat (the chance 
probability is raised by approximately a factor 2), but 
for doublets within 4$^{\circ}$ the chance 
probability remains below 1 \%.

%table5
To examine the effect of the choice of coordinates,
the analysis is repeated for the G coordinate system.
There are no multiplets found, even for the largest 
space angles (within 5$^{\circ}$)
and within $\pm10^{\circ}$ of the G plane.
The average number of doublets expected
by chance is 2.8. 
The Poisson probability of observing no multiplets is thus 6 \%.
Considering that the number of events within $\pm10^{\circ}$ 
of the G plane is
about 30 \% less than that expected from uniformity
(see Fig.2), this probability is statistically reasonable.

\bigskip
The details of each event which are members of one of 
the clusters  within 4$^{\circ}$
are listed in Table \ref{tab:events}.
Here one HP event is counted
twice (doublet \#5 and \#8), since it forms a doublet
with another HP event, and with a VR event.
Clearly some, at least, of these multiplets must arise by chance.
In an attempt to select genuinely coincident events from sources 
at limited distances,  it may be better
to only use events above a slightly higher energy threshold, to avoid
the systematic differences in energy determination of each experiment. 
Selecting only events above $5\times10^{19}$ eV, two triplets and 
one doublet (\#7) 
remain, from a total number of 51 events.
The chance probability of the triplet is now 1.1 \%, falling to 0.1 \%
if we examine only the restricted region within $\pm 10^{\circ}$ of 
the SG plane.

%table6

\bigskip
The order of arrival times with respect to energy
of the events in triplet\#1 are not as expected from a
recent, nearby bursting source where cosmic rays are
accerelated in a short time \cite{sigl98}.
According to simulations on the propagation of
protons through both the intergalactic and Galactic 
magnetic fields by Medina Tanco  \cite{Tanco98},
the extragalactic magnetic field must be 
much smaller than the present upper limit of
10$^{-9}$ G to explain the present clustering
within experimental angular resolution.

In the case of triplet \#2, the direction is 
consistent with the Ursa-Major II cluster of galaxies.
The magnetic field strength in this cluster of
galaxies is possibly of the order of sub-$\mu$G, similar to 
that observed in the Coma cluster \cite{Kim91}.  
If so,
each member of the triplet might be a gamma-ray, because protons
may not be collimated within $2.5^{\circ}$ and the mean distance
of travel before decay for neutrons is 1 Mpc for 10$^{20}$ eV.
Since the cross-section of pair creation by gamma-rays of this energy
and the probability of bremsstrahlung of the resulting electrons
are suppressed due to the relativistic
contraction of the atmosphere (LPM effect \cite{Landau35,Migdal56}), the
longitudinal development of a gamma-ray shower is greatly depressed and
delayed.
A ground array, therefore, might grossly underestimate 
the primary energy of such a shower if it is estimated by
the local particle density around 600m from
the shower core \cite{lawrence91,takeda99}. 
However, if the geomagnetic field component normal to
the arrival direction is large enough, electron-positron pair creation 
in the geomagnetic field occurs far from the
earth and a cascade develops in the stratoshpere\cite{McBreen81}. 
Therefore in the northern hemisphere we might observe
gamma-ray showers in the highest energy region mainly from
a northerly direction \cite{Kasahara96}.
The terrestrial arrival directions of the three members of 
triplet \#2 are indeed from the north, which is at least
consistent with this conjecture.
However, it should be noted that
the attenuation length of gamma-rays of energies above
$4\times10^{19}$ eV due to the interaction on
radio photons is less than 10 Mpc \cite{bhattacharjee98}
and hence a source distance limit also applies for gamma-rays.

If each member of the triplets are protons, coming
from the same source, then the intervening magnetic field must be
so weak that the particles are hardly deflected,
or there is magnetic focusing in the magnetic field 
structure of the Local Supercluster as demonstrated 
by Lemoine et al. \cite{Lemoine99}.

\begin{figure}[H]
\centerline{\epsfile{file=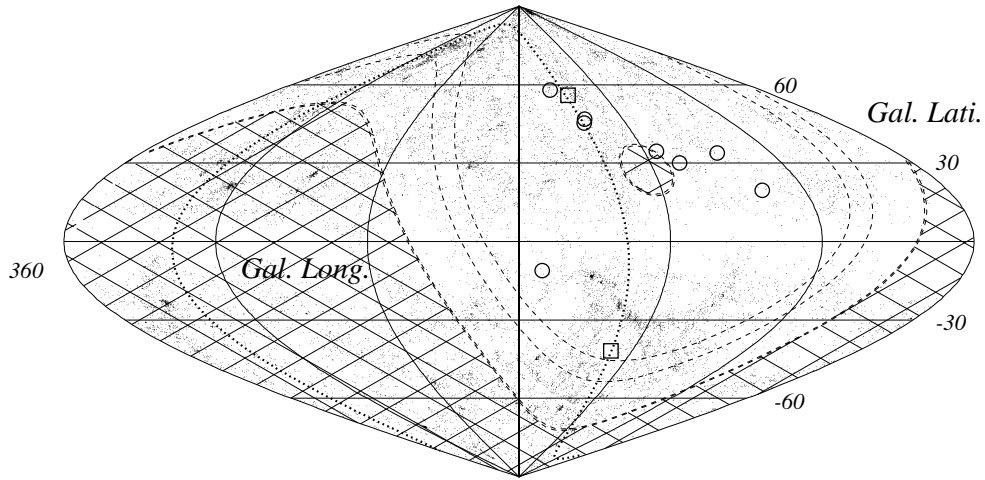,height=7cm}}
  \caption{\it The location of galaxies within 100 Mpc
               in G coordinates from the CfA 1995 catalogue and the
               clusters within $4^{\circ}$.
               The squares show triplets and the circles
               show doublets.}
  \label{fig:cfa}
\end{figure}

\medskip
In Figure \ref{fig:cfa}, the directions of triplets (open squares) and
doublets (open circles) are plotted on the CfA galaxy \cite{cfa}
distribution within 100 Mpc.  There are no multiplets from the most
crowded region (Virgo Cluster) and there seems no correlation
with the density of nearby galaxies.
In the following we look for any source candidate for
the triplets \#1 and \#2, and the doublet \#7.

The AG highest energy event and HP 10$^{20}$ eV event
are members of the triplet \#1 and this triplet may
be related to a nearby source.
Mrk 359 (l= 134.60$^{\circ}$, b=-42.87$^{\circ}$) 
with z=0.017 (68 Mpc assuming $H_0$=75 km/s/Mpc)
is within 2.3$^{\circ}$ from the direction of the centroid of
this triplet.
This direction is also within 3 and 6 degrees of a clusterings of events
with energies above 10$^{19}$ eV claimed by 
Chi et al.  (l$\simeq133^{\circ}$, b$\simeq-40^{\circ}$) \cite{Chi92}
and Efimov and Mikhailov (RA=27$^{\circ}$, Dec=18$^{\circ}$ or
 l=143$^{\circ}$, b=-42$^{\circ}$) \cite{Efimov94}.
Al-Dargazelli et al. \cite{Al-Dar} pointed that the colliding
galaxy pair VV338 (l=138$^{\circ}$, b=-34$^{\circ}$) 
(N672 and U1249) may be related to the clustering around the
region l=135$^{\circ}$ and b=-35$^{\circ}$, where they identified
an apparent clustering above 10$^{19}$ eV.
Though VV338 is very close, 5.7 Mpc,
the angular separation from triplet \#1 is large (9.0$^{\circ}$).

Triplet \#2 comprises three AG events of similar energies,
(5-8)$\times10^{19}$ eV, and is within 2.4$^{\circ}$ from the direction
of the interacting galaxy, VV144 (Mkn40 or Arp151; l=147.03$^{\circ}$,
b=58.54$^{\circ}$).
This is a Seyfert galaxy with z=0.020, corresponding to about 81 Mpc.
Triplet \#2 is on top of a maximum in the arrival direction
probability simulated by Medina Tanco \cite{Tanco98} for sources
located between 20 and 50 Mpc.
These simulations assume that the luminous matter
in the nearby universe tracks the distribution of cosmic ray sources
and modulates the intensity of IGMF.

In the direction of doublet \#7, there are interacting galaxies
VV101 (l=87.06$^{\circ}$, b=33.75$^{\circ}$, 100 Mpc)
with a space angle separation of 1.3$^{\circ}$ 
and VV89 (l=88.14$^{\circ}$, b=35.51$^{\circ}$, 14.5 Mpc)
with a separation of 2.7$^{\circ}$.

From the above discussion, if the extragalactic magnetic field
is much weaker than the present upper limit of 10$^{-9}$G,
then triplets \#1 and \#2 and doublet \#7 are possibly
correlated with interacting galaxies, as pointed out 
by Al-Dargazelli et al \cite{Al-Dar}.
Clearly, the next generation experiments
are needed to confirm such an association.

Recently Lemoine et al. \cite{Lemoine99} made extensive simulations of
the propagation of protons in a large scale magnetic field of 
strength $\sim 0.05-0.5 \mu$G in the Local Supercluster.  
They found an 8-20 \% probability of detecting 5 doublets above 40 EeV 
in the AGASA data set of 47 events, due to magnetic focusing.
It is important to examine what kind of conditions
are required for the magnetic field configuration in the
Local Supercluster to explain the present results of two triplets
along the SG plane.

\bigskip

\section{Conclusion}

A number of collimated triplets and doublets within 4$^{\circ}$
(the approximate angular resolution of combined data set)
are observed.
The chance probability is of the order of 10 \% and is not 
statistically significant.
However, the chance of observing triplets and doublets
within $\pm 10^{\circ}$ of the SG plane is less than 1 \%.
It should be noted that the probability of observing
the multiplets above 5$\times 10^{19}$ eV is about 1 \% and
it is less than 0.1 \% if attention is limited to
within $\pm 10^{\circ}$ of the SG plane.
Though there is still a possibility of chance coincidence,
we should pay more attention on the small angle clustering
along the SG plane
in the interpretation of EHECR enigma.

We expect detailed investigation with much better angular
resolution and much higher statistics from future experiments.
In particular we note that the probabilities given
in this paper are 'a postiori' in that hints of
clustering have been reported in earlier analysis and these data
have been included here.
There are insufficient events for an independent test of
our observations.
An independent set awaits new projects such as Hi-Res, Auger
and Telescope Array.

\section*{Acknowledgement}
We gratefully acknowledge the very considerable efforts of 
the members of the experiments at Volcano Ranch,
Haverah Park, Yakutsk and AGASA.  
We thank
P. Sokolsky for valuable discussion during his stay at Institute
for Cosmic Ray Research, University of Tokyo.

\newpage

\begin{description}
  \item[Fig.1]
      Declination distribution of showers of energies above 10EeV
      and zenith angles smaller than 45$^{\circ}$ of AG, YK and HP.
  \item[Fig.2]
      Galactic (left) and supergalactic latitude (right)
      distribution of arrival directions of 92 cosmic rays
      from four ground array
      experiments.
      Energy thresholds are 40EeV.   Solid lines are the sum of expected
      distributions of each experiment for a uniform distribution.
  \item[Fig.3]
      Sky map of 92 events above 40EeV in equatorial coordinates.
      Squares - AGASA, Triangles - Haverah Park,
      Circles - Yakutsk and Stars - Volcano Ranch.
      The region of sky observable in each experiment is shown by dashed lines
      from the top YK, HP, AG and VR.
  \item[Fig.4]
      Sky map of 92 events above 40EeV in galactic coordinates.
      The symbols are the same as used in figure 3.
  \item[Fig.5]
      Frequency distribution of the number of doublets in 1,000,000
      simulated data sets within several space angles.
      The arrow mark shows the observed number of doublets.
      Chance probability of
      the observed number of doublets is calculated by summing
      the number of data sets in the hatched region.
  \item[Fig.6]
      Arrival directions with error circles of EHECR
      in supergalactic coordinates.
      The total number of events is 92.
  \item[Fig.7]
      Probability of observing doublets and triplets within
      10$^{\circ}$ from the SG plane with 1,000,000 simulated
      data sets for 92 events.
  \item[Fig.8]
      The location of galaxies within 100 Mpc
      in G coordinates from the CfA 1995 catalogue and the
      clusters within $4^{\circ}$.
      The squares show triplets and the circles
      show doublets.
\end{description}

\newpage

\bigskip

\begin{table}
\begin{tabular}[b]{|c||r|r|r|r|r|} \hline
  Experiment & Longitude & Latitude &   Number of &  Zenith  &  Error in
arrival \\
           &           &          &  events &  angle &  determination  \\
\hline \hline
  AGASA & 138$^{\circ}$30'E &  38$^{\circ}$47'N & 47 &  $<45^{\circ}$ &
$1.8^{\circ}$ \\ \hline
  Haverah Park &  1$^{\circ}$38'W & 53$^{\circ}$58'N & 27 &  $<45^{\circ}$ &
$3^{\circ}$  \\ \hline
  Yakutsk & 129$^{\circ}$24'E & 61$^{\circ}$42'N &   12 &  $<45^{\circ}$ &
$3^{\circ}$  \\ \hline
  Volcano Ranch &  106$^{\circ}$47'W & 35$^{\circ}$09'N &  6 &
$<45^{\circ}$ & ($3^{\circ}$) \\ \hline \hline
  Total & & & 92 & & \\ \hline
\end{tabular}
\caption{\it Experimental sites and number of events above 40EeV.}
\label{tab:site}
\end{table}

\renewcommand{\arraystretch}{1.0}

\bigskip
\begin{table}
\begin{center}
\begin{tabular}{|l|r|} \hline
  Combination & Resolution (degrees) \\ \hline \hline
  AG-AG & 2.5 \\ \hline
  AG-YK, AG-HP, AG-VR & 3.5 \\ \hline
  HP-HP, YK-YK, VR-VR & 4.2 \\ \hline
  HP-YK, HP-VR, YK-VR & 4.2 \\ \hline
\end{tabular}
\end{center}
\caption{\it Combined space angle resolution from angular resolution
             of each experiment.}
\label{tab:error}
\end{table}

\bigskip
\begin{table}
\begin{center}
\begin{tabular}{|r||r|r||r|r|} \hline
            & \multicolumn{2}{|c||}{Doublet} & \multicolumn{2}{|c|}{Triplet}
\\ \hline
 \multicolumn{1}{|c||}{space angle} & \multicolumn{1}{|c|}{observed} &
\multicolumn{1}{|c||}{probability} & \multicolumn{1}{|c|}{observed} &
\multicolumn{1}{|c|}{probability}  \\
            & \multicolumn{1}{|c|}{number}   &             &
\multicolumn{1}{|c|}{number}  &              \\ \hline \hline
  $<3.0^{\circ}$ & 12 &  1.5\%  & 2 &  1.4\%  \\ \hline
  $<4.0^{\circ}$ & 14 & 13.4\%  & 2 &  8.3\%  \\ \hline
  $<5.0^{\circ}$ & 20 & 15.9\%  & 3 & 11.8\%  \\ \hline
\end{tabular}
\end{center}
\caption{\it Probability of observing doublets and triplets with 1,000,000
             simulated data sets.  Each triplet is counted as three doublets.}
\label{tab:prob}
\end{table}

\bigskip
\begin{table}
\begin{center}
\begin{tabular}{|r||r|r||r|r|} \hline
            & \multicolumn{2}{|c||}{Doublet} & \multicolumn{2}{|c|}{Triplet}
\\ \hline
 \multicolumn{1}{|c||}{space angle} & \multicolumn{1}{|c|}{observed} &
\multicolumn{1}{|c||}{probability} & \multicolumn{1}{|c|}{observed} &
\multicolumn{1}{|c|}{probability}  \\
            & \multicolumn{1}{|c|}{number}   &             &
\multicolumn{1}{|c|}{number}  &              \\ \hline \hline
  $<3.0^{\circ}$ &  8 &  0.1\% & 2 &  0.2\%  \\ \hline
  $<4.0^{\circ}$ &  9 &  0.3\% & 2 &  0.9\%  \\ \hline
  $<5.0^{\circ}$ & 11 &  0.6\% & 2 &  3.1\%  \\ \hline
\end{tabular}
\end{center}
\caption{\it Probability of observing doublets and triplets within
$\pm$ 10$^{\circ}$ from the SG plane for 92 events.}
\label{tab:prob_sg}
\end{table}

\bigskip
\begin{table}
\begin{center}
\begin{tabular}{|r||r|r||r|r|} \hline
            & \multicolumn{2}{|c||}{Doublet} & \multicolumn{2}{|c|}{Triplet}
\\ \hline
 \multicolumn{1}{|c||}{space angle} & \multicolumn{1}{|c|}{observed} &
\multicolumn{1}{|c||}{probability} & \multicolumn{1}{|c|}{observed} &
\multicolumn{1}{|c|}{probability}  \\
            & \multicolumn{1}{|c|}{number}   &             &
\multicolumn{1}{|c|}{number}  &              \\ \hline \hline
  $<3.0^{\circ}$ &  8 &  0.2\% & 2 &  0.3\%  \\ \hline
  $<4.0^{\circ}$ &  9 &  0.8\% & 2 &  1.7\%  \\ \hline
  $<5.0^{\circ}$ & 11 &  1.5\% & 2 &  5.4\%  \\ \hline
\end{tabular}
\end{center}
\caption{\it Probability of observing doublets and triplets when
  an excess of 20 \% in the $0^{\circ} \sim 30^{\circ}$ range of
  SG latitide is assumed.}
\label{tab:prob_excess}
\end{table}

\bigskip

\begin{table}
\begin{center}
\begin{tabular}{|r||r||r||r||r|r||r|r||r|r|} \hline
 Cluster & Exp. & Date & Log E & R.A. & Dec. & l & b & S.G.Lng. &
 S.G.Lat. \\ \hline \hline
 Triplet \#1 & HP & 810105 & 19.99 & 20.00 & 20.00 & 132.70 & -41.70 &
 318.10 & -0.79 \\
             & AG & 931203 & 20.33 & 18.91 & 21.07 & 130.48 & -41.44 &
 318.11 & 0.89 \\
              & AG & 951029 & 19.71 & 18.53 & 20.03 & 130.18 & -42.51 &
 317.02 & 0.93 \\ \hline
 Triplet \#2  & AG & 920801 & 19.74 & 172.30 & 57.14 & 143.20 & 56.65 &
  56.82 & 2.04 \\
              & AG & 950126 & 19.89 & 168.65 & 57.58 & 145.53 & 55.10 &
  55.51 & 0.51 \\
              & AG & 980404 & 19.73 & 168.44 & 55.99 & 147.51 & 56.23 &
  56.84 & -0.37 \\ \hline  \hline
 Doublet \#1  & AG & 910420 & 19.64 & 284.90 & 47.79 &  77.88 & 18.45 &
  24.95 & 57.83 \\
              & AG & 940706 & 20.03 & 281.36 & 48.32 &  77.58 & 20.86 &
  29.35 & 57.26 \\ \hline
  Doublet \#2  & AG & 860105 & 19.74 &  69.03 & 30.15 & 170.08 & -11.50 &
 350.38 & -33.33 \\
              & AG & 951115 & 19.69 &  70.39 & 29.85 & 171.09 & -10.79 &
 351.23 & -34.31 \\ \hline
Doublet \#3 & HP &  860315 & 19.71 & 267.00 & 77.00 & 108.50 &  30.10 &
  30.83 &  27.99 \\
         & AG & 960513 & 19.68 & 269.05 & 74.12 & 105.11 &  29.79 &
  31.09 &  30.94 \\ \hline
   Doublet \#4  & HP & 720525 & 19.65 & 239.00 & 79.00 & 113.30 &  34.60 &
  35.05 &  23.27 \\
              & YK & 911201 & 19.62 & 235.40 & 79.80 & 114.60 &  34.60 &
  34.88 &  22.22 \\ \hline
 Doublet \#5  & VR & 610319 & 19.73 & 154.10 & 66.70 & 143.00 &  44.30 &
  44.59 &   0.35 \\
              & HP & 850313 & 19.62 & 157.00 & 65.00 & 143.60 &  46.30 &
  46.63 &   0.24 \\ \hline
 Doublet \#6  & HP & 661008 & 19.67 & 164.00 & 50.00 & 159.00 &  58.80 &
  61.08 &  -5.53 \\
              & YK & 750317 & 19.67 & 163.70 & 52.90 & 154.90 &  56.80 &
  58.45 &  -4.16 \\ \hline
 Doublet \#7  & HP & 740228 & 19.86 & 264.00 & 58.00 &  86.36 &  32.52 &
  41.02 &  45.22 \\
              & AG & 980330 & 19.84 & 259.16 & 56.32 &  84.39 &  35.17 &
  45.44 &  45.35 \\ \hline
 Doublet \#8  & HP & 760206 & 19.62 & 165.00 & 64.00 & 140.98 &  49.43 &
  49.49 &   2.41 \\
              & HP & 850313 & 19.62 & 157.00 & 65.00 & 143.60 &  46.30 &
  46.63 &   0.24 \\ \hline
\end{tabular}
\end{center}
\caption{\it Event lists of members of clusters within $4^{\circ}$
             from 4 surface experiments AGASA(AG), Haverah Park(HP),
             Volcano Ranch(VR) and Yaktuk(YK).}
\label{tab:events}
\end{table}

\newpage

\bigskip

\bigskip

\bigskip

\bigskip

\bigskip

\bigskip

\bigskip

\bigskip

\end{document}